\documentclass[showpacs]{revtex4}
\usepackage{amsmath}
\usepackage{amssymb}
\usepackage{graphicx}

\begin{document}

\title{
Chameleon stars supported by a cosmological scalar field
 }

\author{
Vladimir Folomeev
\footnote{Email: vfolomeev@mail.ru}
}
\affiliation{
Institute of Physicotechnical Problems and Material Science of the NAS
of the
Kyrgyz Republic, 265 a, Chui Street, Bishkek, 720071,  Kyrgyzstan \\
}
\begin{abstract}
Starting from the assumption that the present accelerated expansion of the Universe is
driven by a chameleon scalar field, the function
describing a direct coupling  between the matter content of the Universe
and the scalar field is derived
from  cosmological considerations.
In progressing from cosmology towards astrophysical scales, this function is used to estimate the influence
that such nonminimal coupling may have on the properties of usual (polytropic) stars
whose matter can interact directly with the cosmological chameleon.
It is shown that in order to obtain sizes and masses of compact configurations
supported by the chameleon scalar field comparable to those of
usual stars,
it is necessary to assume that the pressure of the matter content of the Universe
is substantially different from zero.
The question of a possible manifestation of the effects of the nonminimal coupling in laboratory experiments
is also discussed.
\end{abstract}

\maketitle

\section{Introduction}

The numerous attempts to give a theoretical description of the current accelerated expansion of the Universe are usually reduced
to introducing
an exotic form of matter called dark energy, which is believed to be responsible for such an acceleration.
Whereas  the true nature of dark energy is still unknown,
different possibilities are being suggested to describe it.
Perhaps one of the most promising
approaches to addressing the origin of dark energy is a consideration of
 theories that include
 various fundamental fields \cite{sahni:2004,Copeland:2006wr}.
 Possibilities are also being considered, including some that involve
   modified (non-Einstein) gravity theories \cite{DeFelice:2010aj,Nojiri:2010wj} or models with extra space dimensions
\cite{Maartens:2003tw,Dzhunushaliev:2009va}.
 In any case, a description of the present acceleration of the Universe within the framework of the aforementioned  approaches implies
 that one needs to choose
 the
 parameters entering these theories in such a way as to satisfy the current
 astronomical observations and laboratory experiments.

One of such approaches is the so-called
 chameleon paradigm, the essence of which, briefly,  is as follows~\cite{Khoury:2003aq,Khoury:2003rn}:
Based on the assumption that  scalar fields with a very small mass of order $H_0 \sim 10^{-33}~\text{eV}$
do exist in the Universe, and they drive the current accelerated expansion,
it is assumed that such fields may
 interact directly with  ordinary matter and dark matter
  with gravitational strength. This leads to the fact that  the scalar field
acquires a mass which depends on the local background matter density.
 For example, on Earth, where the density is high, the mass of the field becomes
 $10^{30}$ times greater than the mass of this field on the cosmological background.

An important ingredient  of the chameleon mechanism is the presence of two functions  of the scalar field
$\varphi$: the potential energy
$V(\varphi)$ and the function $f(\varphi)$ describing the direct coupling  between
the matter and the scalar field. These two functions are arbitrary, and they are chosen
in such a way as to satisfy various laboratory and cosmological tests
\cite{Khoury:2003aq,Khoury:2003rn,Brax:2004qh,Mota:2006fz}.

It is obvious that this arbitrariness in the choice of the mentioned field functions is
due to lack of knowledge of
the true nature of scalar fields filling the Universe
(of course, assuming they do really exist in the present Universe and are responsible for its
accelerated expansion).
Perhaps the only thing one can do at this point is to try to find the restrictions
imposed upon
the form
of these functions and  values of the parameters appearing in them.
Some work has already been
done along these lines, particularly
 in the papers
 \cite{Farajollahi:2010pk, Cannata:2010qd,Chattopadhyay:2011fp}, where
the authors choose a generalized expression
for the nonminimal coupling between the scalar field and the matter having the form $f(\varphi) L_m$,
with $L_m$ being the Lagrangian of ordinary matter.
In considering the  evolution of the Universe they have used
two different approaches:
in Ref.~\cite{Farajollahi:2010pk} the functions $V(\varphi)$ and  $f(\varphi)$
are taken to be  arbitrary, and their parameters are chosen in such a way as to satisfy the current observational data.
Another approach has been applied in
 \cite{Cannata:2010qd,Chattopadhyay:2011fp}, where the authors have initially selected some particular  form of cosmological evolution and
 found the functions  $V(\varphi)$ and  $f(\varphi)$ corresponding to such evolution.

On the other hand, the question may be asked: if the mechanism of direct coupling between
 a chameleon scalar field and ordinary matter is universal, i.e., it can be applied to any form of matter
 (be it ordinary matter or dark matter),
 how will this affect the properties
 of various
small-scale objects such as, for instance, usual stars?
In Einstein gravity, compact configurations consisting of ordinary matter and scalar fields have been repeatedly considered
earlier in the literature.
The scalar fields may be self-interacting or not, and one can
involve a fluid which interacts with the scalar field either only gravitationally \cite{Henriques:1989ar,Henriques:1989ez}
or through  direct coupling as well \cite{Lee:1986tr,Crawford:2009gx}.
 In particular,
in the recent papers \cite{Dzhunushaliev:2011ma,Folomeev:2011uj,Folomeev:2011aa}
the changes in the structure of compact objects (polytropic
stars) under the influence of a chameleon field have been investigated.
It was shown that, depending on the properties of {\it arbitrarily } chosen functions
$V(\varphi)$ and  $f(\varphi)$,
the properties of these configurations, such as total mass, distribution of matter, and size, depend strongly
on the surrounding scalar field.

The purpose of the present paper is to demonstrate the possibility of determining
 the nonminimal coupling function $f(\varphi)$ which
satisfies the current astronomical observations.
Subsequently, this function will be used in modeling compact objects in order to clarify the
influence of the presence of the
 chameleon scalar field on their physical characteristics.
As an example, a polytropic star embedded in the chameleon scalar field
will be considered.

The paper is organized as follows:
In Sec.~\ref{f_cosm} the nonminimal coupling function $f(\varphi)$ is obtained, in general form, from  cosmological
considerations. In the particular case of the $\Lambda$CDM model
this function is derived
in Sec.~\ref{f_particular} and, in Sec.~\ref{cham_star}, is applied
 to the estimation of the influence
of the nonminimal coupling upon the properties of usual stars whose matter can interact directly with a
 chameleon scalar field.
In  the Appendix, 
we consider the effects related to the nonminimal coupling in the sense
of their possible manifestation
in laboratory experiments.
 Finally, in Sec.~\ref{concl_f_cosm} we summarize and discuss the results obtained.

\section{$f$ from cosmology}
\label{f_cosm}

As  pointed out in the Introduction, here the choice of the nonminimal coupling function
 $f(\varphi)$ will be done based on a consideration of the cosmological evolution of the Universe
 within the  chameleon paradigm suggested in Ref.~\cite{Farajollahi:2010pk}.
 To do this,
we choose the Lagrangian of the system in the form
\begin{equation}
\label{lagran_gen}
L=-\frac{c^4}{16\pi G}R+\frac{\Delta}{2}\partial_{\mu}\varphi\partial^{\mu}\varphi -V(\varphi)+f(\varphi) L_m ~.
\end{equation}
Here $\varphi$ is the real scalar field with the potential $V(\varphi)$;
$\Delta=\pm 1$ corresponds to the usual or phantom scalar field, respectively;
$L_m$ is the Lagrangian of ordinary matter
(a perfect fluid). The case $f=1$ corresponds to no direct coupling between
the fluid and scalar field, but even in this case the two sources are still coupled via gravity.

\subsection{About the Lagrangian $L_m$}
\label{about_L_m}

The Lagrangian $L_m$ can be chosen in a variety of ways. For example, in Ref.~\cite{Hawking1973} the choice
 $L_m=-2\varepsilon$ is used,
and in
Refs.~\cite{Stanuk1964,Stanuk}, $L_m=p$ ($\varepsilon$ and $p$ are the energy density and the pressure of a fluid, respectively).
By varying both these  matter Lagrangians with respect to a metric, one obtains the same energy-momentum tensor
of the perfect
fluid in the conventional form. However, it can be shown that, for instance, for static configurations, these Lagrangians,
being substituted  in the general  Lagrangian \eqref{lagran_gen}, will give different equations for an equilibrium
configuration (the Tolman-Oppenheimer-Volkoff (TOV) equations). In the case of
 $L_m=p$, the TOV equation will have the same form as when $f=1$. On the other hand, the use of
$L_m=-2\varepsilon$ leads to the appearance
of an extra term on the right-hand side of the TOV equation
associated with the nonminimal coupling.

Of course, in the limit $f\to 1$, for both choices of $L_m$,
the gravitating system of a  scalar field plus ordinary matter
 reduces   to the same uncoupled system in which the ordinary matter and the scalar field interact only gravitationally.
But in the general case of $f=f(\varphi)$, these two systems will not be equivalent, giving equilibrium configurations
with different properties.
Apparently, at the present time, it is difficult to make a motivated choice amongst these Lagrangians
 $L_m$, or any other Lagrangians  used in the literature
(for the other possible Lagrangians, see,
 e.g., Ref.~\cite{Bertolami:2008ab}). For this reason, a description of various systems with nonminimal coupling of the type
 $f(\varphi) L_m$ is made, in effect,
using an ad hoc choice of  $L_m$.

\subsection{Deriving $f$ in general form}

In this paper we choose  $L_m=p$ which has been used in Refs.~\cite{Dzhunushaliev:2011ma,Folomeev:2011uj,Folomeev:2011aa}
in modeling astrophysical objects.
Here
this choice of  $L_m$ will be used
in studying the cosmological evolution of the present Universe.

Consider the flat model with the metric
\begin{equation}
\label{metr_cosm}
ds^2=c^2 dt^2-a^2(t)dl^2,
\end{equation}
where $a(t)$ is the  scale factor.
Using the general Lagrangian \eqref{lagran_gen}, the corresponding energy-momentum tensor is (details are given in the appendix of
\cite{Dzhunushaliev:2011ma})
 \begin{equation}
\label{emt_cham_star}
T_i^k=f\left[(\varepsilon+p)u_i u^k-\delta_i^k p\right]+\Delta\,\partial_{i}\varphi\partial^{k}\varphi
-\delta_i^k\left[\frac{\Delta}{2}\,\partial_{\mu}\varphi\partial^{\mu}\varphi-V(\varphi)\right]~,
\end{equation}
where $\varepsilon$ and $p$ are the  energy density and the pressure of the fluid, and $u^i$ is the 4-velocity.
Using the metric \eqref{metr_cosm} and the energy-momentum tensor \eqref{emt_cham_star},
the $(_0^0)$ component of the Einstein equations
is then given by
\begin{equation}
\label{Einstein-00_cosm}
\left(\frac{\dot{a}}{a}\right)^2=\frac{8\pi G}{3 c^2}\left[f\varepsilon+\frac{\Delta}{2 c^2}\,\dot{\varphi}^2
+V(\varphi)\right].
\end{equation}
Here the dot denotes differentiation with respect to the cosmological time $t$.
The equation for the scalar field $\varphi$ follows from the Lagrangian \eqref{lagran_gen}
 \begin{equation}
\label{sf_eq_gen}
\frac{1}{\sqrt{-g}}\frac{\partial}{\partial x^i}\left[\sqrt{-g}g^{ik}\frac{\partial \varphi}{\partial x^k}\right]=
\Delta\left(-\frac{d V}{d \varphi}+L_m \frac{d f}{d \varphi}\right)
\end{equation}
and gives in the metric \eqref{metr_cosm}
\begin{equation}
\label{sf_cosm}
\ddot{\varphi}+3\frac{\dot{a}}{a}\dot{\varphi}=
\Delta\, c^2\left(-\frac{d V}{d \varphi}+p\frac{d f}{d\varphi}\right)~.
\end{equation}
One more equation can be obtained from the
law of conservation of energy and momentum,
$T^k_{i;k}=0$. Taking the $i=0$ component of this equation gives
\begin{equation}
\label{conserv_cosm}
\frac{d(f\varepsilon)}{dt}+ 3\frac{\dot{a}}{a} f(\varepsilon+p)+p \frac{d f}{dt}=0.
\end{equation}
Keeping in mind
that $\varepsilon$ and $p$ are related by some equation of state,  there
are three unknown functions: $a, \varphi, \varepsilon$, and three equations
\eqref{Einstein-00_cosm}, \eqref{sf_cosm}, and \eqref{conserv_cosm} for them.

Now we need to choose an equation of state for the matter content of the Universe.
At the present time, it is believed  that the main part of matter in the Universe (except dark energy)
is in the form of dark matter. The most popular hypothesis in modeling dark matter
is the assumption that its equation of state corresponds to cold (dust) matter \cite{sahni:2004}.
Such matter is either pressureless or has a very small pressure (see, e.g., Ref.~\cite{Muller:2004yb}).
In hydrodynamical language, in cosmology, the relation between the pressure and energy density
is usually taken in the simplest form as $p=w\varepsilon$, where $w=\text{const}$.
Assuming $w\approx 0$ at the present time, a description of the evolution of the Universe is carried out.

Here we initially start with the assumption that the pressure of the matter
filling the Universe is small compared with its energy density.
To apply the results obtained from cosmology for modeling compact astrophysical objects,
let us choose a polytropic equation of state which is
widely used in astrophysics. This equation of state may be taken in the following parametric form:
\begin{equation}
\label{eq_cosm_star}
p=K \rho^{1+1/n}, \quad \varepsilon=\rho c^2.
\end{equation}
Here the constant $n$ is called the polytropic index and is related to the specific heat ratio $\gamma$ via
$n=1/(\gamma-1)$, $\rho$ is the rest-mass density,
and $K$ is  an arbitrary constant
whose value depends on the properties of the fluid under consideration.
Substituting this equation of state in
 \eqref{conserv_cosm}, its general solution will be
\begin{equation}
\label{conserv_sol_gen}
\frac{\rho}{(1+\alpha \rho^{1/n})^n}=\frac{\rho_0}{f a^3}, \quad \alpha=K/c^2,
\end{equation}
where $\rho_0$ is an integration constant.
The smallness of the pressure implies that the term in the denominator
$\alpha \rho^{1/n} \ll 1$. Then, neglecting this term, we obtain from Eq.
\eqref{conserv_sol_gen} the following approximate expression for the mass density:
\begin{equation}
\label{conserv_sol_appr}
\rho\approx \frac{\rho_0}{f a^3}~.
\end{equation}
In the limit $f\to 1$, i.e., in the absence of the nonminimal coupling, this expression reduces to the usual
dependence for the density of dust matter as a function of the scale factor.

For the subsequent calculations, it is convenient to
introduce new dimensionless variables
\begin{equation}
\label{dimless_var}
\xi=B t, \quad B=\sqrt{8\pi G \rho_0}, \quad \phi(\xi)=\left(\frac{8\pi G}{c^4}\right)^{1/2}\varphi(t),
\end{equation}
where $B$ has the dimensions of inverse time.
Using these variables and taking into account the equation of state
 \eqref{eq_cosm_star}  and the expression \eqref{conserv_sol_appr},
 the system of Eqs.~\eqref{Einstein-00_cosm} and \eqref{sf_cosm} takes the form
\begin{eqnarray}
\label{Einstein-00_cosm_dmls}
&&3\left(\frac{a^\prime}{a}\right)^2=\frac{1}{a^3}+\frac{\Delta}{2}\phi^{\prime 2}+\tilde{V},
 \\
\label{sf_cosm_dmls}
&&\phi^{\prime\prime}+3 \frac{a^\prime}{a} \phi^{\prime}=
\Delta\left(-\frac{d\tilde{V}}{d\phi}+\frac{\sigma}{f^\gamma a^{3\gamma}}\frac{d f}{d\phi}\right).
\end{eqnarray}
Here the prime denotes differentiation with respect to the dimensionless time $\xi$;
 $\gamma=1+1/n$, $\tilde{V}=V/(\rho_0 c^2)$,
$\sigma=p_0/(\rho_0 c^2)=K \rho_0^{1/n}/c^2$, where $\rho_0, p_0$ refer to the current values of the mass density and the pressure,
respectively.
Note that in our low-pressure approximation the parameter
 $\sigma\ll 1$. Also, in this approximation, the first term on the right-hand side of Eq.~\eqref{Einstein-00_cosm_dmls}
always plays the role of dustlike matter, independently  of the form of the function
$f$. In turn, the latter, together with
 $V$, will describe an evolution of all other forms of matter, including dark energy.

Next, using Eqs.~\eqref{Einstein-00_cosm_dmls} and \eqref{sf_cosm_dmls},
let us find the form of the function
 $f$ expressed in terms of the scale factor
 $a$. To do this, rewrite Eq.~\eqref{sf_cosm_dmls} as
\begin{equation}
\label{sf_cosm_dmls_2}
\frac{1}{2}\frac{d(\phi^{\prime 2})}{d\xi}+3 \frac{a^\prime}{a} \phi^{\prime 2}=
\Delta\left(-\frac{d\tilde{V}}{d\xi}+\frac{\sigma}{f^\gamma a^{3\gamma}}\frac{d f}{d\xi}\right).
\end{equation}
Expressing from
 \eqref{Einstein-00_cosm_dmls}
\begin{equation}
\label{sf_deriv}
\phi^{\prime 2}=2\Delta\left[3\left(\frac{a^\prime}{a}\right)^2-\frac{1}{a^3}-\tilde{V}\right]
\end{equation}
and substituting this in
 \eqref{sf_cosm_dmls_2}, one can find the following dependence of
 $f$ on $a$:
\begin{equation}
\label{f_a_depend}
f^{1-\gamma}=6\frac{1-\gamma}{\sigma}\int
 a^{\prime} a^{3\gamma-4}\left(a^2 a^{\prime\prime}+2 a a^{\prime 2}-\frac{1}{2}-a^3 \tilde{V}\right)d\xi.
\end{equation}
Then, if   $a$ and $\tilde{V}$ are known as functions of the time
$\xi$, it is possible to find the corresponding dependence  $f=f(\xi)$.

\subsection{Expression for
 $f$ in the case of the $\Lambda$CDM model}
 \label{f_particular}

We now will demonstrate how to obtain an expression for
 $f$ in the particular case of
$\tilde{V}=0$ (a massless scalar field), and when the ansatz for the scale factor
$a$ is chosen in the form of the well-known $\Lambda$CDM model \cite{sahni:2004}
\begin{equation}
\label{ansatz_a}
a(\xi)=a_0 \left(\frac{\Omega_m}{\Omega_{\Lambda}}\right)^{1/3} \left(\sinh{\frac{\sqrt{3}}{2}}
\sqrt{\frac{\Omega_{\Lambda}}{\Omega_{m}}}\,\xi\right)^{2/3}.
\end{equation}
Here $\Omega_{m}=8\pi G \rho_{0}/3 H_0^2$, $\Omega_{\Lambda}=c^2 \Lambda/3 H_0^2$, where $H_0=H(t=t_0)$
and $a_0=a(t=t_0)$
are the present values of the Hubble parameter and the scale factor, respectively. This expression for $a$
 smoothly interpolates between a matter-dominated Universe in the past
($a \propto t^{2/3}$) and accelerated expansion in the future ($a \propto \exp{\sqrt{\Lambda/3}\,c t}$).

Setting  $a_0=1$ at the present epoch and substituting
 \eqref{ansatz_a} in \eqref{f_a_depend}, one can find
\begin{equation}
\label{f_xi_depend}
f^{1-\gamma}=2\frac{1-\gamma}{\sigma \gamma}\left(\frac{\Omega_{\Lambda}}{\Omega_{m}}\right)^{1-\gamma}
\left(\sinh{\frac{\sqrt{3}}{2}}
\sqrt{\frac{\Omega_{\Lambda}}{\Omega_{m}}}\,\xi\right)^{2\gamma}.
\end{equation}
In turn,  we have from \eqref{sf_deriv}
\begin{equation}
\label{phi_xi_depend}
\phi^{\prime 2}=2\Delta \frac{\Omega_{\Lambda}}{\Omega_m} \quad \Rightarrow \quad
\phi=\sqrt{2\frac{\Omega_{\Lambda}}{\Omega_m}}\,\xi \quad \text{and} \quad
\xi=\frac{1}{\sqrt{2}}\sqrt{\frac{\Omega_{m}}{\Omega_{\Lambda}}}\,\phi.
\end{equation}
(Here we have set  $\Delta=+1$ in order to ensure that the functions $f$ and $a$ obtained below will be real.)
Note here that by substituting the above derivative of the scalar field into Eq.~\eqref{Einstein-00_cosm_dmls} 
and multiplying it by
 $\Omega_m$, we get the known equation for
the $\Lambda$CDM model. That is, in the case under consideration, the gradient of the scalar field   plays the role of
the $\Lambda$-term.

Substituting the expression for
$\xi$ from \eqref{phi_xi_depend} in Eqs.~\eqref{ansatz_a} and \eqref{f_xi_depend}, we find the final formulas for
 $f$ and $a$ as functions of the scalar field
$\phi$
\begin{eqnarray}
\label{a_phi_depend}
&&a=\left(\frac{\Omega_m}{\Omega_{\Lambda}}\right)^{1/3} \left(\sinh{\sqrt{\frac{3}{8}}\,\phi}\right)^{2/3},
 \\
\label{f_phi_depend}
&&f=\frac{\Omega_{\Lambda}}{\Omega_m}\left[-\frac{\sigma(n+1)}{2}\right]^n
\left(\sinh{\sqrt{\frac{3}{8}}\,\phi}\right)^{-2(n+1)}.
\end{eqnarray}

Using  \eqref{a_phi_depend}, one can find the current value of the scalar field,
 $\phi_0\equiv \phi(t=t_0)$. To do this, setting
 $a=a_0=1$ and choosing, for definiteness,
 $\Omega_{\Lambda}=0.7$ and $\Omega_{m}=0.3$,  from the solution of the transcendental equation
\eqref{a_phi_depend}, we have
\begin{equation}
\label{phi_xi_cosm}
\phi_0 \approx 1.975, \quad \xi_0 \approx 0.915.
\end{equation}
This $\xi_0$ corresponds to $t_0\approx 12.95$ Gyrs.

\section{Chameleon star}
\label{cham_star}

We now will proceed to a consideration of a compact configuration (a star) consisting of ordinary (polytropic) matter
with the equation of state  \eqref{eq_cosm_star} and of a scalar field nonminimally interacting with this matter.
In Ref.~\cite{Dzhunushaliev:2011ma} this type of gravitating configuration
of a polytropic fluid nonminimally coupled to a scalar field was called a ``chameleon star.''

As a nonminimal coupling function, we will use the expression \eqref{f_phi_depend} obtained in the preceding  section.
We proceed from the assumption that the cosmological chameleon scalar field
penetrates into the star and
interacts with
the star's matter through the same universal coupling function  $f$ from
\eqref{f_phi_depend}. That is, we assume  that
the matter of the star, which is embedded in the external, cosmological, homogeneous scalar field,
feels its presence not only gravitationally,     but also through the nonminimal coupling.
Then, the structure of the star is also defined by the distribution of the scalar field,
and correspondingly by $f$, inside the star.

\subsection{General equations for hydrostatic equilibrium configurations}

We choose the static spherically symmetric metric in the form
\begin{equation}
\label{metric_sphera}
ds^2=e^{\nu} c^2 dt^2-e^{\lambda}dr^2-r^2d\Omega^2,
\end{equation}
where $\nu$ and $\lambda$ are functions of the radial coordinate $r$,
and $d\Omega^2$ is the metric on the unit two-sphere.
The $(_0^0)$ and $(_1^1)$  components of the Einstein equations for the metric
\eqref{metric_sphera} and the energy-momentum tensor \eqref{emt_cham_star} are then given by
\begin{eqnarray}
\label{Einstein-00_cham_star}
&&G_0^0=-e^{-\lambda}\left(\frac{1}{r^2}-\frac{\lambda^\prime}{r}\right)+\frac{1}{r^2}
=\frac{8\pi G }{c^4}\left(f \varepsilon +\frac{1}{2}\,e^{-\lambda} \varphi^{\prime 2}+V\right),
 \\
\label{Einstein-11_cham_star}
&&G_1^1=-e^{-\lambda}\left(\frac{1}{r^2}+\frac{\nu^\prime}{r}\right)+\frac{1}{r^2}
=\frac{8\pi G }{c^4}\left(-f p -\frac{1}{2}\,e^{-\lambda} \varphi^{\prime 2}+V\right),
\end{eqnarray}
where the prime denotes differentiation with respect to $r$.
(As in the previous section, we have again set   $\Delta=+1$.)
Here the energy density $\varepsilon$ and the pressure $p$ refer now to the matter of the star
(but not to the cosmological matter), but the scalar field, in the spirit of the
chameleon paradigm, remains the same as in cosmology.
Our purpose will be to clarify
the influence that this {\it cosmological} scalar field may have on the inner structure and characteristics of such
compact objects as usual stars.

Equations \eqref{Einstein-00_cham_star} and \eqref{Einstein-11_cham_star} must be supplemented by an equation
for the scalar field, which follows from
 \eqref{sf_eq_gen} and has the following form in the metric \eqref{metric_sphera}:
\begin{equation}
\label{sf_cham_star}
\varphi^{\prime\prime}+\left[\frac{2}{r}+\frac{1}{2}\left(\nu^\prime-\lambda^\prime\right)\right]\varphi^\prime=
e^{\lambda}\left(\frac{d V}{d \varphi}-p\frac{d f}{d\varphi}\right)~.
\end{equation}
 One more equation follows
from the
law of conservation of energy and momentum,
$T^k_{i;k}=0$. Taking the $i=1$ component of this equation gives
\begin{equation}
\label{conserv_2_cham_star}
\frac{d p}{d r}=-\frac{1}{2}(\varepsilon+p)\frac{d\nu}{d r}.
\end{equation}

To perform the subsequent calculations, it is convenient to rewrite the equation of state
 \eqref{eq_cosm_star} in terms of
the new variable
$\theta$ related to the density $\rho$ at the given point and the central density $\rho_c$ by
(see, e.g., Ref.~\cite{Zeld})
\begin{equation}
\label{theta_def}
\rho=\rho_c \theta^{n_s} \quad \Rightarrow \quad p=K\,\rho^{\gamma_s}=K \rho^{1+1/n_s}=K\rho_c^{1+1/n_s} \theta^{n_s+1}.
\end{equation}
Notice that $\theta$ must be ``normalized'' to unity when $r=0$.
Using \eqref{theta_def} in Eq.~\eqref{conserv_2_cham_star} leads to
\begin{equation}
\label{conserv_3_cham_star}
2\sigma_s (n_s+1)\frac{d\theta}{d r}=-(1+\sigma_s \theta)\frac{d\nu}{dr},
\end{equation}
with $\sigma_s=K \rho_c^{1/n_s}/c^2=p_c/(\rho_c c^2)$ and $p_c$ is the pressure of the fluid at the center of the configuration.
The index $s$ (``star'') on $\sigma, n$, and $\gamma$ is introduced here to distinguish these
 parameters from those used in Sec.~\ref{f_cosm} in describing
the matter content of the Universe. In both cases, the relativistic parameters
 $\sigma$ and $\sigma_s$ characterize the ratio of the pressure at a given point
(in the case of a  star, at the center of the configuration),
or at a given instant of time (in the case of cosmology, at the present time) to the energy density.
Since the expression for the coupling function  $f$ from \eqref{f_phi_depend} has been obtained
in the low-pressure approximation,
i.e., is valid only when
 $\sigma \ll 1$, therefore, it is natural that in the case of a star this expression is also applicable
 only when  $\sigma_s \ll 1$.

Equation \eqref{conserv_3_cham_star} may be integrated to give $e^{\nu}$ in terms of $\theta$
\begin{equation}
\label{nu_app}
	e^{\nu}=e^{\nu_c}\left(\frac{1+\sigma_s}{1+\sigma_s \theta}\right)^{2(n_s+1)},
\end{equation}
where $e^{\nu_c}$ is the value of $e^{\nu}$ at the center of the configuration where $\theta_c=1$.
The integration constant $\nu_c$ corresponds to the value of $\nu$ at the center of the
configuration. It is determined by requiring $e^{\nu}=1$ at infinity, i.e., that the
space-time be asymptotically flat.

We introduce a new function $M(r)$
\begin{equation}
\label{u_app}
e^{-\lambda}=1-\frac{2 G M(r) }{c^2 r}~.
\end{equation}
$M(r)$ is the mass of the configuration in the range $0 \leq r \leq r_1$,
where $r_1$ is the boundary of the fluid where $\theta=0$. Also, we introduce
dimensionless variables
\begin{equation}
\label{dimless_xi_v}
x=A r, \quad v(x)=\frac{A^3 M(r)}{4\pi \rho_{c}},
\quad \phi(x)=\left[\frac{4\pi G}{\sigma_s(n_s+1)c^4}\right]^{1/2}\varphi(r),
\quad \text{where} \quad A=\left[\frac{4\pi G \rho_{c}}{\sigma_s (n_s+1) c^2}\right]^{1/2},
\end{equation}
where $A$ has the dimensions of inverse length.
 With this, we can rewrite  Eqs.~\eqref{Einstein-00_cham_star} and \eqref{Einstein-11_cham_star} in the form
 \begin{eqnarray}
\label{eq_v_app}
\frac{d v}{d x} &=& x^2\left\{f\theta^{n_s}+\frac{1}{2}\left[
1-2\sigma_s(n_s+1)\frac{v}{x}
\right]
\left(\frac{d\phi}{d x}\right)^2+\tilde{V}\right\}
,\\
\label{eq_theta_app}
x^2\frac{1-2\sigma_s(n_s+1)v/x}{1+\sigma_s\theta}\frac{d\theta}{d x} &=&
x^3\left[f\theta^{n_s}\left(1-\sigma_s\theta\right)+
2 \tilde{V}-\frac{1}{x^2}\frac{d v}{d x}\right]-v ~,
\end{eqnarray}
where $\tilde{V}=V/(\rho_c c^2)$ is the dimensionless potential energy of the field.
Next, using  \eqref{conserv_3_cham_star}, one can rewrite Eq.~\eqref{sf_cham_star} as follows:
\begin{eqnarray}
\label{eq_phi_dim_cham_star}
\frac{d^2 \phi}{d x^2}&+&\left\{\frac{2}{x}-\frac{\sigma_s(n_s+1)}{1+\sigma_s \theta}
\left[\frac{d\theta}{d x}+\frac{1+\sigma_s \theta}{1-2\sigma_s(n_s+1)v/x}\,\frac{1}{x}
\left(\frac{d v}{d x}-\frac{v}{x}\right)\right]\right\}\frac{d\phi}{d x} \nonumber\\
&&=\left[1-2\sigma_s(n_s+1)\frac{v}{x}\right]^{-1}\left(\frac{d \tilde{V}}{d\phi}-\sigma_s \theta^{n_s+1}\frac{d f}{d\phi}\right).
\end{eqnarray}
Thus, the static configuration under consideration is described by the three equations \eqref{eq_v_app}-\eqref{eq_phi_dim_cham_star}.

\subsection{The particular choice of
 $f$ in the form \eqref{f_phi_depend}}

We now apply Eqs.~\eqref{eq_v_app}-\eqref{eq_phi_dim_cham_star} for describing the inner structure of
our configuration. In Ref.~\cite{Dzhunushaliev:2011ma}, this system was used in studying similar configurations
(chameleon stars) when the coupling function $f$ was chosen arbitrarily.
It was shown  that
 the presence of the nonminimal interaction leads to substantial changes both in the radial matter distribution of the stars
and in the star's total mass.

To demonstrate the influence of the nonminimal coupling derived in Sec.~\ref{f_cosm}
on the properties of a compact configuration,
here we will consider a case when the potential energy is absent, i.e., $V(\phi)=0$.
As mentioned above, we assume that the scalar field, being initially homogeneous (i.e., cosmological),
in the presence of star's matter, begins to interact with the matter both gravitationally and through the
nonminimal coupling.
Our purpose will be to study the possible effects of this interaction.

Note that, in the absence of the nonminimal coupling, the choice $V(\phi)=0$ corresponds to the case of a massless scalar field.
However,
one can see from the right-hand side of Eq.~\eqref{eq_phi_dim_cham_star}
that the presence of the  nonminimal coupling leads to the presence of an additional term
of the form $\sigma_s \theta^{n_s+1} d f/d\phi$.

Choosing  $f$ in the form  \eqref{f_phi_depend}, Eqs.~\eqref{eq_v_app}-\eqref{eq_phi_dim_cham_star}
are to be solved for given $\sigma_s$, $n_s$, and also cosmological parameters $\sigma$, $n$, $\Omega_\Lambda$, and $\Omega_m$
 subject to the boundary conditions in the vicinity
of the center of the configuration $x=0$,
\begin{equation}
\label{bound_all}
\theta \approx \theta_c+\frac{\theta_2}{2}\, x^2, \quad v \approx v_3 x^3, \quad
\phi \approx \phi_c+\frac{\phi_2}{2}\,x^2,
\end{equation}
where  $\phi_c$ corresponds to the central value of the scalar field
$\phi$, $\theta_c$ is normalized to be unity at the center, $\theta_c\equiv\theta(x=0)=1$.
The values of the coefficients
$\phi_2$, $v_3$, and $\theta_2$ are defined from Eqs.~\eqref{eq_phi_dim_cham_star}, \eqref{eq_v_app}, and \eqref{eq_theta_app},
respectively, as
$$
\phi_2=-\frac{1}{3}\, \sigma_s \left(\frac{d f}{d \phi}\right)_c,
\quad v_3=f_c/3, \quad
\theta_2 \approx -\frac{f_c}{3}(1+4\sigma_s),
$$
where the index $c$
corresponds  to central values.

Since the cosmological scalar field,  $\phi(\xi)$, and that used in modeling a star,
$\phi(x)$, are measured in different units, then, to find the relation between them,
we compare their definitions
 given by  expressions \eqref{dimless_var} and \eqref{dimless_xi_v}, respectively. Thus we have
\begin{equation}
\label{relation_phis}
\phi(\xi)=\sqrt{2\sigma_s(n_s+1)}\,\phi(x).
\end{equation}
Substituting this expression in
 \eqref{f_phi_depend}, we finally obtain
\begin{equation}
\label{f_star_final}
f=\frac{\Omega_{\Lambda}}{\Omega_m}\left[-\frac{\sigma(n+1)}{2}\right]^n
\left(\sinh{\sqrt{\frac{3}{4}\,\sigma_s(n_s+1)}\,\phi(x)}\right)^{-2(n+1)}.
\end{equation}

As a result, the static configuration under consideration is described by the system of 
Eqs.~\eqref{eq_v_app}-\eqref{eq_phi_dim_cham_star} together with the boundary conditions
\eqref{bound_all} and the function
 $f$ from \eqref{f_star_final}.
 These equations, however, cannot be integrated analytically, and have to be solved numerically.
The solutions were started near   the
 vicinity
of the center of the configuration
(i.e., near $x\approx 0$) and solved out to a point
$x = x_1$,  where the function $\theta$ became zero. Since our spherical fluid configuration
was taken to be embedded in an external, homogeneously distributed scalar field, we also needed to require that at the
boundary, $x = x_1$, the value of the varying scalar field from the inside of the star,
$\phi (x)$,
matched the exterior value $\phi_{\text{ext}}$. In this case, of course, the gradient of the scalar field, generally speaking,
although small, is not strictly equal to zero.
It requires continuing the solutions outside the fluid.
Formally, the gradient goes to zero only asymptotically, i.e., as
 $x\to \infty$. However, the numerical calculations indicate that for small values of the parameters
 $\sigma$ and $\sigma_s$ used in this paper,
the gradient of the scalar field  is virtually
 equal to zero along the radius of the configuration.
Then the form of the external scalar field is determined by matching it to the interior solution at the surface of
the star. The junction conditions, which follow from the requirement that
the field equations be satisfied on the boundary of the star, require that $\phi (x)$ be continuous.
In our case of small scalar-field gradients,
this can be done with sufficient
accuracy  by equating the interior and exterior values of the scalar field at $x = x_1$,
i.e., by requiring that  $\phi(x_1) \approx \phi_{\text{ext}}$.

In turn, for a complete description of the system under consideration,
it is also necessary to solve external Einstein's equations
for the metric functions $\nu$ and $\lambda$ from \eqref{metric_sphera}.
This can be easily done in the manner described in Ref.~\cite{Folomeev:2011aa}.
However, for the problems being considered in this paper,
knowledge
of the behavior of the exterior solutions is not necessarily needed.
That is why, in order not to encumber the paper,  we do not show this procedure here.


Thus, the procedure for finding solutions is as follows: for given values of $\sigma$, $\sigma_s$,
$\Omega_{\Lambda}$, $\Omega_{m}$, $n$,
and $n_s$, we choose
the central value of the scalar field, $\phi_c$, such that $\theta$ goes to zero and
$\phi(x_1) \approx \phi_{\text{ext}}$ at some finite value of $x = x_1$.
The fact that the configuration is embedded in the external cosmological scalar field
$\phi_0$ implies  that
$\phi_{\text{ext}}$ must be equal to
$\phi_0/\sqrt{2\sigma_s (n_s+1)}$, where $\phi_0$ is  taken
from \eqref{phi_xi_cosm} [the coefficient $\sqrt{2\sigma_s (n_s+1)}$ appears because of the relation
\eqref{relation_phis}]. This allows one to determine definitely the value of
$\phi_c$ for such configurations.

One can see from the expression \eqref{f_star_final}
that the function $f$ can be positive, negative, or imaginary depending on the value of $n$.
Physically reasonable values of $f$ are either positive or negative.
However, the numerical calculations show that the parameters  used in the paper give
finite size solutions only for positive values of this function. To provide the positivity  of
$f$, it is necessary to take $n$ either even
  or equal to
\begin{equation}
\label{relat_n}
 n=\frac{2 l}{2 m+1},
\end{equation}
 where  $m$ and $l$ are integers. Proceeding from this choice, we get the results described below.

In order not to encumber the paper,
in further calculations the polytropic indexes $n$ and $n_s$ are taken to be equal to each other.
The calculations indicate that even if this is not the case,
the qualitative
behavior of the solutions remains the same. Keeping this in mind,
we present here the results of numerical calculations with
the following values of the parameters $n$ and $n_s$:
$n=n_s=6/5; 2; 8/3$ (for convenience, we hereafter drop the index  $s$ on $n_s$).
These values cover the range
 $1.2< n \lesssim 2.67$, which is often used in modeling usual stars. The parameters
 $\Omega_{\Lambda}$ and $\Omega_{m}$ are chosen   to be
$\Omega_{\Lambda}=0.7$, $\Omega_{m}=0.3$, as before.
In the approximation used here, the relativistic
parameters $\sigma, \sigma_s$ are taken to be much less than unity: $\sigma \lesssim 0.1$, $\sigma_s \lesssim 0.05$.

\begin{figure}[t]
\centering
  \includegraphics[height=9cm]{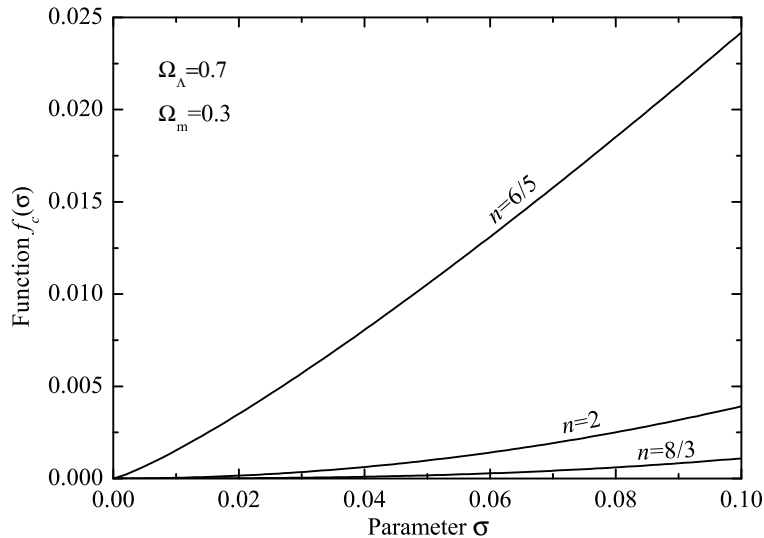}
\vspace{-1.cm}
\caption{The function $f_c$ given by Eq.~\eqref{f_sigma_n_approx} versus $\sigma$ for $n=6/5, 2, 8/3$.}
\label{fig_f_sigma}
\end{figure}

With this choice of the parameters, the numerical calculations indicate that the scalar field
$\phi(x)$ and the coupling function $f$ are virtually constant
 along the radius of the
configuration under consideration. This allows us to rewrite the system
\eqref{eq_v_app}-\eqref{eq_phi_dim_cham_star} with $f$ from \eqref{f_star_final},  using the
approximation $\phi^{\prime} \approx 0$ and
$f(x)\approx f_c=\text{const}$, in the following form:
\begin{eqnarray}
\label{eq_v_approx}
\frac{d v}{d x} &=& f_c x^2  \theta^n,\\
\label{eq_theta_approx}
x^2\frac{1-2\sigma_s(n+1)v/x}{1+\sigma_s\theta}\frac{d\theta}{d x} &=&
x^3\left[f_c \theta^n\left(1-\sigma_s\theta\right)
-\frac{1}{x^2}\frac{d v}{d x}\right]-v~.
\end{eqnarray}
Then, introducing the new dimensionless radial coordinate $x=\tilde{x}/\sqrt{f_c}$ and
mass function $v=\tilde{v}/\sqrt{f_c}$,
we are able to get rid of $f_c$ in these equations. Thus, mathematically this system of equations becomes equivalent to
that of describing usual polytropic stars with an equation of state in the form
 \eqref{eq_cosm_star} and in the absence of a scalar field. Such configurations were investigated in detail
 in the work of Tooper \cite{Tooper:1964} who
 found regular,
relativistic starlike configurations. These starlike solutions of \cite{Tooper:1964}
were found for the values of the parameters $0 \leq \sigma_s \leq 0.75$  and $1 \leq n \leq 3$.

In our case we are dealing with approximately the same range of $n$, but with small values of
 $\sigma_s$. Correspondingly, the size and the mass of our configurations
are $1/\sqrt{f_c}$ times the values of those  considered in \cite{Tooper:1964}
in the absence of a scalar field.
 That is, in our approximation the value of $f_c$ is the primary
 determining parameter of the problem.

It is obvious that $f_c$ must depend on the values of three parameters: $n$, $\sigma$, and $\sigma_s$.
However, the numerical calculations show that, for fixed  $n$ and varying
$\sigma$ and $\sigma_s$, the argument of the
hyperbolic sine in \eqref{f_star_final} is a slowly varying function of $\sigma_s$.
That is why, to a good approximation, this argument may be regarded as a constant,
that allows to rewrite the function $f_c(\sigma)$  in the following approximate form:
\begin{equation}
\label{f_sigma_n_approx}
f_c(\sigma)\approx \bar{f}_n \left[-\frac{\sigma(n+1)}{2}\right]^n, \quad \text{with} \quad
\bar{f}_n=\frac{\Omega_{\Lambda}}{\Omega_m}\left(\sinh{\sqrt{\frac{3}{4}\,\sigma_s(n+1)}\,\phi(x)}\right)^{-2(n+1)}
\approx \text{const},
\end{equation}
where $\bar{f}_n $ is approximately constant along the radius of the configuration for any fixed $n$.
For the values of  $n$  used here, the coefficient
$\bar{f}_n$ has the following numerical values: $\bar{f}_{n=6/5} \approx 0.342$, $\bar{f}_{n=2} \approx 0.174$, and
$\bar{f}_{n=8/3} \approx 0.101$. The numerical calculations indicate that,  using
these values,
the approximation for $f_c(\sigma)$ in the form  \eqref{f_sigma_n_approx} is applicable with good accuracy
when $\sigma \lesssim 0.1$ and $\sigma_s \lesssim 0.05$.

Figure~\ref{fig_f_sigma} shows the function  $f_c(\sigma)$ from \eqref{f_sigma_n_approx}.
One can see from this figure that the value of $f_c(\sigma)$, being always much less than unity,
grows with increasing  $\sigma$ for all $n$.
Since the size and the mass of the configurations under consideration are  inversely proportional to
the square root of  $f_c(\sigma)$, then its small values for any
 $\sigma$ and $n$ correspond to large sizes and masses compared with configurations without a scalar field
 considered in Ref.~\cite{Tooper:1964}.
The numerical results obtained in \cite{Tooper:1964} are written in terms of dimensionless variables
in the form \eqref{dimless_var}
for arbitrary values of the equation-of-state parameters $K$, $n$, and the central density
 $\rho_c$. Correspondingly, the concrete values of physical characteristics (masses and sizes) can be obtained in
 dimensional form by giving the corresponding values of the mentioned parameters.
But in any case,
the size and the mass of configurations with a scalar field considered in the present paper
increase catastrophically in the limit of very small
$\sigma$, corresponding to  dustlike cosmological matter.

\section{Summary and discussion}
\label{concl_f_cosm}

It is now believed that various scalar fields may exist in the present Universe.
Based on the assumption that   they can interact directly (nonminimally)  with  ordinary matter and dark matter
  with gravitational strength, this paper studies the possible influence that such fields may have
on the properties of compact gravitating configurations (stars)
consisting of ordinary matter. The choice of the nonminimal coupling function $f$
is made from a consideration of the evolution of the Universe within the framework
of  chameleon cosmology. Then the total Lagrangian initially contains a term describing a
nonminimal coupling between the scalar field $\varphi$ and the matter in the general form
 $f(\varphi) L_m$, with $L_m$ being the Lagrangian of ordinary matter [see Eq. \eqref{lagran_gen}].
By choosing the arbitrary function $f(\varphi)$,
one might try to describe the evolution of the present Universe.

In the present paper the choice of this function is based on two propositions:
(i) The pressure of the matter filling the Universe is small compared with its energy density.
(ii) The evolution of the cosmological scale factor
$a$ is determined by
the observational data.

To model the matter content of the Universe (namely, ordinary matter plus dark matter), we choose a polytropic equation of state
in the form  \eqref{eq_cosm_star}. This choice is caused by our wish to use the results obtained from cosmology
in modeling compact objects (see below). For this equation of state, in the low-pressure approximation,
it is possible to find a general expression for $f$ as a function of
$a$ in the form \eqref{f_a_depend}. Next, using the well-known $\Lambda$CDM ansatz for the scale factor,
we find $f$ as a function of the scalar field in the final form \eqref{f_phi_depend}.

Our next step is to consider the influence that the obtained nonminimal coupling  may have
on the properties of  usual stars. While so doing, we start from the assumption that a star's matter is described
by the same equation of state \eqref{eq_cosm_star} and also can interact directly with a cosmological scalar field.
One can consider
such a configuration as embedded in the scalar field which
penetrates into the star and  interacts with its matter through the function $f$.

Depending on a concrete form of $f$, such a direct coupling results in more or less significant changes in the
structure and properties of a star (see Refs.~\cite{Dzhunushaliev:2011ma,Folomeev:2011uj,Folomeev:2011aa}).
In our low-pressure approximation, when the pressures both of the cosmological matter and of the star's fluid
are much less than their energy densities, the function $f$, as well as the scalar field,
may be regarded as practically constant
along the radius of the star: $f(x)\approx f_c=\text{const}$. This allows one to rewrite the system of
Eqs.~\eqref{eq_v_app}-\eqref{eq_phi_dim_cham_star}, describing a static configuration, in the form which is
mathematically identical  to that  without a scalar field
[see Eqs.~\eqref{eq_v_approx} and \eqref{eq_theta_approx}].
This corresponds to the fact that the size and the mass of our configurations
are $1/\sqrt{f_c}$ times the values obtained in Ref.~\cite{Tooper:1964}, where
configurations with the same equation of state, but
without a scalar field, have been considered.
But since in our case the values of $f_c$ are always much less than unity, the size and the mass of configurations with the
chameleon scalar field are always much greater
 than those of configurations without a scalar field.
It is seen from Fig.~\ref{fig_f_sigma} that this effect is especially strong for small values of the cosmological
parameter  $\sigma$ which correspond to the vanishing pressure of the matter filling the Universe
(dark matter plus ordinary matter). This  gives rise to the fact that
the size and the mass of compact configurations with the chameleon scalar field increase
catastrophically compared with configurations without a scalar field.

Does this unusual behavior of the characteristics of the compact configurations
considered here inevitably imply that such chameleon cosmology is nonviable?
The answer to this question might be given in the following way:
As one can see from Fig.~\ref{fig_f_sigma},
$f_c$ increases with increasing  $\sigma$,
and correspondingly the size and the mass of our configurations decrease.
One might expect  that the subsequent growth of  $\sigma$,
when the low-pressure approximation is not already valid, and the function
 $f$ is not constant throughout the star, will provide  the possibility of obtaining
physically reasonable sizes and masses
of configurations supported by the chameleon scalar field. At least
the previous studies with an arbitrary choice of
$f$
performed in Refs.~\cite{Dzhunushaliev:2011ma,Folomeev:2011uj,Folomeev:2011aa}
indicate that one can obtain sizes and  masses
comparable to those of usual stars without a scalar field.
However, such a problem with  $f$ derived from cosmological considerations must be studied separately
beyond the framework of the low-pressure approximation used in the present paper.

Thus, within the framework of chameleon cosmology considered here,
polytropic stars supported by the cosmological chameleon scalar field
will have physically reasonable sizes and masses only if we
suppose that the pressure of
the matter filling the Universe is nonvanishing.
However, in the case of relatively small pressures
considered here, this does not lead to any significant changes from the point of view
of conventional cosmology. Indeed, it is seen from the cosmological Einstein equation \eqref{Einstein-00_cosm_dmls}
that its right-hand side is identical in form to the  source of matter in the form of  dust plus the terms
containing the scalar field which describe dark energy, regardless of its form.
In the particular case of the $\Lambda$CDM ansatz for the scale factor considered here,
the massless scalar field  gave the contribution which is  equivalent to the
$\Lambda$-term.

In the case when pressures both of the matter content of the Universe
and of star's matter are not small compared with their energy densities, one would expect that
the situation  will change.
On the one hand, one can expect to get more reasonable sizes and masses of compact configurations.
On the other hand, from the point of view of cosmology, matter sources of the Einstein equations
would be expected generally to have a different
form from that of usually used in modeling the evolution of the Universe.
But these questions should be considered in further studies.

We have also addressed the issue of
how the
effects related to the nonminimal coupling
could be manifested in laboratory experiments.
One  such effect is the occurrence  of extra accelerations whose values
must lie within the restrictions imposed by various experiments.
Proceeding from a consideration of the equation of motion for a fluid element
in a given gravitational field,
in the Appendix the restriction imposed by the E\"{o}tv\"{o}s-type experiments is derived [see Eq.~\eqref{accel_restr}].
Despite the fact that the nonminimal function $f$ obtained here leads to
the unusual behavior of the characteristics of compact configurations, we use this function
to demonstrate the theoretical possibility of
how one can estimate extra accelerations which appear in the Earth's atmosphere because of the
presence of the nonminimal coupling.
It is shown that the condition \eqref{accel_restr} is satisfied in principle
[see Eq.~\eqref{accel_restr_earth}].
In future work, we plan to perform a consideration of these questions using such a nonminimal coupling function,
which gives a more adequate description of the characteristics of compact configurations.

\section*{Acknowledgements}
The author is grateful to the Research Group Linkage
Programme of the Alexander von Humboldt Foundation for the
support of this research.
This work was partially supported by a grant in fundamental research in natural
sciences by the Ministry of Education and Science of Kazakhstan.

\appendix
\numberwithin{equation}{section}

\boldmath
\section{
Effects of the nonminimal coupling and laboratory constraints
}
\unboldmath
\label{WEP}

In this appendix we address the question of how the effects
related to the presence of the nonminimal coupling in the system described by the Lagrangian
\eqref{lagran_gen} could manifest themselves in laboratory experiments.

To illustrate this, consider the equation of motion for a fluid element
in a gravitational field.
We derive this equation using
the energy-momentum tensor
 \eqref{emt_cham_star} which can be rewritten  in the form
\begin{equation}
\label{emt_cham_star_geod}
T^{\mu \nu}=f \,T^{\mu\nu \text{(f)}}+T^{\mu\nu \text{(sf)}},
\end{equation}
where
$$
T^{\mu\nu \text{(f)}}=(\varepsilon+p)u^{\mu} u^{\nu}-g^{\mu \nu} p, \quad
T^{\mu\nu \text{(sf)}}=
\Delta\,\partial^{\mu}\varphi \partial^{\nu}\varphi
-g^{\mu \nu}\left[\frac{\Delta}{2}\,\partial_{i}\varphi\partial^{i}\varphi-V(\varphi)\right]~.
$$
Here the total energy-momentum tensor is divided into two parts denoted by
the indexes $\text{(f)}$ and $\text{(sf)}$ which refer to the fluid and the scalar field, respectively.
On operating a covariant derivative on \eqref{emt_cham_star_geod}, we have
$$
T^{\mu \nu}_{;\,\nu}=f \,T^{\mu\nu \text{(f)}}_{;\,\nu}+T^{\mu\nu \text{(f)}} f_{;\,\nu}+T^{\mu\nu \text{(sf)}}_{;\,\nu}.
$$
By virtue of the scalar field equation
 \eqref{sf_eq_gen}, the third term on the right-hand side of
 this expression vanishes.
Projecting two remaining terms
 onto the direction of the fluid 4-velocity $u_{\mu}$, one can obtain
the following equation of motion for a fluid element:
\begin{equation}
\label{eq_geod}
\frac{d^2 x^{\mu}}{d s^2}+\Gamma^{\mu}_{\alpha \nu}\frac{d x^{\alpha}}{d s}\frac{d x^{\nu}}{d s}=
-\frac{p_{;\,\nu}+p\,(\ln{f})_{;\,\nu}}{\varepsilon+p}\left(u^{\mu}u^{\nu}-g^{\mu\nu}\right).
\end{equation}
The first term in the numerator on the right-hand side, $p_{;\,\nu}$, is the pressure gradient that also appears
in usual non-chameleon models and describes the force exerted on a fluid element due
to the fluid pressure (it is not to be attributed to the coupling between the matter and
the scalar field, nor does it signal any new effect).
The presence of this term forces
particles of matter, within which there are pressures that generally move along lines that are not geodesics.
The second term in the numerator, $p\,(\ln{f})_{;\,\nu}$, is provided by the presence of the nonminimal coupling
in the system. One can see that its influence on the motion of a test particle is determined both by the
magnitude of the fluid pressure and the gradient of the nonminimal
coupling function $f$. In the general  case when  magnitudes of the pressure and the gradient of $f$
 are not small, one might expect  that substantial extra accelerations in the system described by the Lagrangian
 \eqref{lagran_gen} will appear.

The presence of such extra accelerations
implies that the theory may exhibit
violations of the weak equivalence principle.
The most stringent constraints on possible violations of the equivalence principle derive from experiments performed on Earth.
These experiments test
the universality of free fall, the prediction that
all bodies in a uniform gravitational field have exactly
the same gravitational acceleration. The universality of free fall has been
tested to roughly 1 part in $10^{12}$ \cite{Zeld} using laboratory test
bodies in the gravitational fields of the Earth and the Sun.
Another sort of experiments is connected with precise measurements
 by using lunar laser-ranging  to compare the accelerations of the Earth and Moon toward
the Sun. These experiments constraint the difference in free-fall acceleration
 to be less than approximately one part in $10^{13}$ (see, e.g., Ref.~\cite{Baessler:1999iv}).

Then, proceeding from the requirement to satisfy the aforementioned laboratory experiments,
one must make sure
that the effects related to the presence of the nonminimal coupling do not lead to a contradiction with
 the experiments.
To check this, we will estimate below the value of extra acceleration which appears in the presence
of the nonminimal coupling with the function $f$ given by the expression \eqref{f_phi_depend}.
But before doing that, it will be useful to notice that  since we consider here the case of low-pressure approximation,
when  $f(x)\approx f_c=\text{const}$,  the gradient of the function  $f$ is virtually
 equal to zero along the radius of the configuration.
 This  allows us to expect
 that the influence
 of the term
$p\,(\ln{f})_{;\,\nu}$ in
Eq.~\eqref{eq_geod} will be
 small compared with the  gradient term
 $p_{;\,\nu}$. Indeed, the numerical calculations show that  the
smaller the value of the relativistic parameter $\sigma_s$, the smaller the value of the term
 $p\,(\ln{f})_{;\,\nu}$ compared with  $p_{;\,\nu}$.
 One would expect that in the nonrelativistic limit, when $\sigma_s\to 0$,
 the extra acceleration associated with the nonminimal coupling  will be negligibly
small compared with the  acceleration provided by a gravitational field.

It is natural to assume that the nonrelativistic approximation used in the present paper is also relevant
in describing relatively small objects
such as planets. Consistent with this, let us estimate the value of  extra acceleration in a nonrelativistic
(``Newtonian'') system in the presence of a nonminimal coupling such as
 $p f(\varphi)$. In doing so, we will compare the gravitational acceleration of a fluid element with the extra acceleration
 associated with the nonminimal coupling.
To do this, let us find from Eq.~\eqref{eq_geod} the expression for the usual  three-acceleration in the nonrelativistic
approximation
$p\ll \varepsilon$
\begin{equation}
\label{accel_cham}
a_{\text{tot}}\approx a_N\left(1+\frac{a_{\text{ch}}}{a_N}\right),
\end{equation}
where $a_{\text{tot}}$ is the total radial acceleration defined as a sum
of the Newtonian acceleration,
$a_N=-\partial\psi/\partial r$ ($\psi$ is the Newtonian gravitational potential), and
the extra acceleration provided by the chameleon, $a_{\text{ch}}=-(p/\rho)(d\ln{f}/d r) $.

To compare the expression \eqref{accel_cham} with  experimental data,
it is convenient to use   the E\"{o}tv\"{o}s parameter,
$$
\eta\equiv 2\frac{|a_1-a_2|}{|a_1+a_2|},
$$
which describes
the difference in relative free-fall acceleration for two bodies of different composition having
the accelerations $a_1$ and $a_2$.
As mentioned above, $\eta$ is experimentally restricted to be less than $10^{-13}$.
Taking into account the smallness of $\eta$, one can find from the above expression:
\begin{equation}
\label{accel_exper}
a_1\approx a_2 (1+\eta).
\end{equation}

Compare the expression \eqref{accel_cham} with \eqref{accel_exper}: whereas the former gives the magnitude  of the total
acceleration of a fluid element in the presence of the nonminimal coupling, the latter  describes the difference between
the accelerations of  two test bodies  which are manufactured to have different compositions. But in both cases the  
difference in accelerations  is estimated, related to either  the nonminimal coupling (in the first case) or a
different composition of bodies (in the second case).
It is clear that independently of the fact which case is under consideration,  the experiment requires that
the difference in accelerations be less than
 $\eta \approx 10^{-13}$. Then, by identifying formally
 $a_{\text{tot}}=a_1$,
$a_N=a_2$, and $a_{\text{ch}}/a_N=\eta$, we have the following restriction on the extra acceleration from
\eqref{accel_cham}:
\begin{equation}
\label{accel_restr}
|a_{\text{ch}}| \equiv \frac{p}{\rho}\,\frac{d\ln{f}}{d r} \lesssim 10^{-13} a_N.
\end{equation}
Here the pressure $p$ and the mass density $\rho$ refer to  a fluid element
falling onto an attractor.
The interaction of this matter
with a chameleon field through the function $f$  results in the extra acceleration
$a_{\text{ch}}$, the value of which must satisfy
the condition \eqref{accel_restr} to be consistent with the experiment.

It is seen that the condition  \eqref{accel_restr} is trivially satisfied if a continuous medium,
located in the Earth's gravitational field, is pressureless. Such a case corresponds,
for instance, to dustlike matter. If the pressure is nonzero, the problem
requires a separate analysis for each particular case. Then, from the point of view of experiment,
each element of such a fluid can be considered as a test particle moving in a given gravitational field.
In this case the magnitude of extra acceleration will depend on the composition of this medium,
which, in turn, is determined by use of various equations of state
$p=p(\rho)$ for different forms of matter. Then, aside from the fact that  the
extra acceleration does take place for one type of matter,
there will also be additional differences in accelerations for fluids having
different compositions.

\subsection{
Extra accelerations in the Earth's atmosphere
}

To illustrate the above considerations, let us now demonstrate how one can estimate the value of extra accelerations, which could
 appear, for instance,
in the Earth's atmosphere.
As is known, the Earth's atmosphere is not
homogeneous and the air density decreases with altitude; the dependence of
the density $\rho$ on altitude $R$ can be approximated by the barometric formula
\begin{equation}
\label{bar_dens}
\rho=\rho_0 e^{-(R-R_{\bigoplus})/\Delta}.
\end{equation}
 Here $\rho_0$ is the density at sea level, $R=R_{\bigoplus}$, and $\Delta$
is the so-called
scale height of the standard atmosphere, which at the Earth's surface is
approximately equal to 8.5 km.
For our purposes, we need only to pick this layer of the atmosphere, i.e., in further calculations we  assume that
$R_{\bigoplus} \leq R \lesssim (R_{\bigoplus}+\Delta)$. This choice of
 $\rho$ implies that the air temperature does not change with altitude. This allows one to
 describe the atmosphere by using the equation of state for an
 isothermal fluid given in the form
\begin{equation}
\label{eos_fluid}
p=K\rho
\end{equation}
with $K=p_0/\rho_0=c_a^2$, where $c_a$ is the speed of sound in the air, and
$p_0$ is the pressure at sea level.
Using this equation of state and introducing the dimensionless variables
\begin{equation}
\label{dmls_vars}
e^{-\chi}=\frac{\rho}{\rho_0}, \quad
\zeta=\frac{R}{L},\quad
\phi(\zeta)=\frac{\sqrt{4\pi G}}{K}\, \varphi(r),
\end{equation}
where $L=\sqrt{K/(4\pi G \rho_0)}$ has dimensions of length,
one can obtain from the Lagrangian \eqref{lagran_gen}
the following  nonrelativistic equations describing the  isothermal fluid in the presence of a
chameleon scalar field \cite{Folomeev:2011uj}:
\begin{eqnarray}
\label{conserv_2_cham_star_2_dmls}
&&\frac{1}{\zeta^2}\frac{d}{d\zeta}\left(\zeta^2\frac{d\chi}{d\zeta}\right)=f e^{-\chi},
\\
\label{sf_cham_star_dmls}
&&\frac{1}{\zeta^2}\frac{d}{d\zeta}\left(\zeta^2\frac{d\phi}{d\zeta}\right)=-e^{-\chi}\frac{d f}{d\phi}.
\end{eqnarray}
The first equation is the equation of hydrostatic equilibrium of a fluid in a Newtonian gravitational field.
Substituting the given distribution of density
\eqref{bar_dens} in \eqref{conserv_2_cham_star_2_dmls}, one can find
$$
f[\phi(\zeta)]=2 \beta \frac{\exp{[\beta (\zeta-\zeta_{\bigoplus})]}}{\zeta},
$$
where $\beta=L/\Delta \approx 10^3$. Using this expression in \eqref{sf_cham_star_dmls} and
taking into account that $\beta \gg 1$, one can obtain an equation for the scalar field in the form
$$
\frac{1}{2}(\phi^{\prime 2})^{\prime}+\frac{2}{\zeta}\,\phi^{\prime 2}=-\frac{2\beta^2}{\zeta},
$$
where the prime denotes differentiation with respect to $\zeta$. The first integral of this equation is
$$
\phi^{\prime 2}=\frac{C_1}{\zeta^4}-\beta^2.
$$
The value of the integration constant $C_1$ is determined from the condition that at the edge of the atmosphere
 $\phi^{\prime} \approx 0$ (since outside the atmosphere the scalar field makes a smooth transition  into the
 cosmological chameleon field).
Representing
 $\zeta$ in a form $\zeta=\zeta_{\bigoplus}+\delta$, where $\delta=\Delta/L \approx 10^{-3}$,
and substituting this in the above expression, we finally get
\begin{equation}
\label{eq_phi_earth}
\phi^{\prime 2}=
\beta^2\left[\left(\frac{\zeta_{\bigoplus}}{\zeta}\right)^4\left(1+\frac{4\delta}{\zeta_{\bigoplus}}\right)-1\right].
\end{equation}
The solution of this equation gives the distribution of the scalar field in the Earth's atmosphere. We will seek
a solution starting from the Earth's surface, $\zeta=\zeta_{\bigoplus}$,
and choosing the initial value of
 $\phi(\zeta_{\bigoplus})$ in such a way as to get approximately the cosmological value of the field at the outer boundary of the
 atmosphere.

Since the cosmological scalar field,  $\phi(\xi)$, and that used here,
$\phi(\zeta)$, are measured in different units, then, to find the relation between them,
compare their definitions
 given by the expressions \eqref{dimless_var} and \eqref{dmls_vars}, respectively. Thus we have
\begin{equation}
\label{relation_phis_earth}
\phi(\xi)=\sqrt{2}\,\frac{K}{c^2}\,\phi(\zeta).
\end{equation}
Substituting this expression in
 \eqref{f_phi_depend}, we get
\begin{equation}
\label{f_earth_final}
f=\frac{\Omega_{\Lambda}}{\Omega_m}\left[-\frac{\sigma(n+1)}{2}\right]^n
\left(\sinh{\sqrt{\frac{3}{4}}\,\frac{K}{c^2}\,\phi(\zeta)}\right)^{-2(n+1)}.
\end{equation}

Next, the experimental restriction from \eqref{accel_restr} on the extra acceleration near the Earth's surface can be rewritten,
by using the  dimensionless variables from \eqref{dmls_vars}, as follows:
\begin{equation}
\label{accel_restr_earth}
\left[\frac{d\ln{f}}{d \zeta}\right]_{\bigoplus} \lesssim 10^{-10}.
\end{equation}
Here we have used the following numerical values:
 $a_N \approx 10\, \text{m s}^{-2}$ and $K=p_0/\rho_0\approx 10^5 \,\text{m}^2 \text{s}^{-2}$.
Then,  evaluating  $\phi(\zeta)$ numerically by using Eq.~\eqref{eq_phi_earth}, and
substituting this in
 \eqref{f_earth_final}, it must be checked
 whether the condition
\eqref{accel_restr_earth} is satisfied or not. The calculations indicate that the value of the logarithmic derivative of
 \eqref{f_earth_final},
$$
\frac{d\ln{f}}{d \zeta}=-\sqrt{3}\,(n+1) \,\frac{K}{c^2}\,\phi^{\prime}(\zeta)
\coth{\left[\sqrt{\frac{3}{4}}\,\frac{K}{c^2}\,\phi(\zeta)\right]},
$$
near the Earth's surface is of order $10^{-10}$ for the values of $n$ used in the present paper.
Thus,
the condition
\eqref{accel_restr_earth} may be considered as satisfied, in principle.
The estimate for the condition \eqref{accel_restr}
for other forms of matter requires a separate analysis.

\end{document}